# Semantic issues in model-driven management of information system interoperability

**Frederick Benaben[1], Nicolas Boissel-Dallier[1&2], Herve Pingaud[1], Jean-Pierre Lorre[2]**

[1] *Centre de Génie Industriel, Ecole des Mines d'Albi, Albi, France*
[2] *PetalsLink, Toulouse France*
{frederick.benaben, nicolas.boissel-dallier, herve.pingaud}@mines-albi.fr
{nicolas.boissel-dallier, jean-pierre.lorre}@petalslink.com

**ABSTRACT:** The MISE Project (Mediation Information System Engineering) aims at providing collaborating organizations with a Mediation Information System (MIS) in charge of supporting interoperability of a collaborative network. MISE proposes an overall MIS design method according to a model-driven approach, based on model transformations. This MIS is in charge of managing (i) information, (ii) functions and (iii) processes among the information systems (IS) of partner organizations involved in the network. Semantic issues are accompanying this triple objective: How to deal with information reconciliation? How to ensure the matching between business functions and technical services? How to identify workflows among business processes? This article aims first, at presenting the MISE approach, second at defining the semantic gaps along the MISE approach and third at describing some past, current and future research works that deal with these issues. Finally and as a conclusion, the very "design-oriented" previous considerations are confronted with "run-time" requirements.

## 1 Introduction

Organisations (including enterprises, associations, institutions, etc.) are strongly dependent from their ability to successfully manage collaborations and to assume the involved interoperability functions: exchange of information, coordination of functions and orchestration of processes. Furthermore, inside such an organisation, Information Systems (IS) and computed systems are assuming both the roles of interface (external and internal exchanges) and functional engine (driving processes and business activities). Therefore, considering that crucial position of IS and computed systems, the previously listed interoperability functions must be supported by these IS. The issue is to ensure that the IS of the partners involved in the collaboration will be able to work altogether (thanks to these interoperability functions) in order to constitute a coherent and homogeneous set of IS (the IS of the collaborative situation).

The MISE project (Mediation Information System Engineering), presented in (Benaben, 2008) and (Benaben, 2010), aims at defining a model-driven design approach of a MIS (Mediation Information System), in charge of interoperability functions. This MIS could be considered as a third party, based on Service-Oriented Architecture (SOA) principles, allowing existing IS to work fluently altogether, according to a common behaviour, without any special effort. This goal is fully compliant with the definition of interoperability given in (Konstantas, 2005) and (Pingaud, 2009) and the interoperability framework given in (Panetto, 2008). The MIS design framework defined in MISE, dealing with business requirements in order to deploy a technical mediation solution, should necessarily succeed in the business-to-technical alignment: the abstract solution (proposed at the business level) must be precisely implemented by the concrete solution (deployed at the technical level). To reach this goal, the semantic attributes must be appropriately defined at the abstract level and rigorously taken into account at the concrete level.

Following the previous considerations, the main objective of this article is to present and discuss the semantic issues embedded into the MISE project, according to two horizontal levels: abstract (business) and concrete (technology) among three vertical layers: informational (information vs. data), functional (activity vs. service) and behavioural (process vs. workflow). Semantic reconciliation is the core of this article.

The second section of this article introduces the MISE project and the associated general principles. The third section identifies semantics issues inherent in that approach. The fourth section presents the current state of the art concerning that kind of semantic problems and the way they can be treated. The fifth section presents specific solutions that have been applied during the first iteration of MISE project (and especially in a French funded project based on MISE results: ISyCri – Interoperability of Systems in Crisis situations). The sixth section concerns the second iteration of MISE and outlines the current research works dealing with these semantic issues. The seventh section concludes this article by enlarging the semantic consideration to non-functional requirements and run-time (while the current article mainly focuses on functional considerations an design-time).

## 2. Overview of the MISE project

The following global presentation of the MISE project will be based on three main parts: (i) overall big picture of the design approach, (ii) model transformation principles and (iii) detailed presentation of each step of the design approach.

### 2.1 Overall big picture of the design approach

This design approach might be seen as a dive into abstraction layers. Consequently it is based on model-driven engineering and on the associated model transformation concepts (OMG, 2003). The general principle is therefore structured according to two steps between three levels:

1. The first step concerns the transition from the "characterization of the situation" level to the "collaborative process models" level. By gathering a structured knowledge concerning the considered collaboration (partners, roles, goals, abilities, etc.) a specific ontology is instantiated to draw a global characterization of the collaborative situation, as described in (Mu, 2011). Then, by applying deduction rules on this knowledge, collaborative processes models are deduced, as described in (Touzi, 2009).
2. The second step concerns the transition from the "collaborative process models" level to the "MIS deployment" level. The knowledge embedded in these collaborative processes models is semantically analysed in order to apply model transformation mechanisms dedicated to match business components (such as business activities from the "collaborative process models" level) with technical components (such as web-services from the "MIS deployment" level). The obtained service-oriented MIS structure, as described in (Benaben, 2010) and (Rajsiri, 2009) can be deployed on the technological target platform, which is an Enterprise Service Bus (ESB). An ESB is a middleware able to efficiently carry message between connected services and, by extension, potentially able to orchestrate workflows between connected services (if a workflow engine is plugged in order to exploit the communication facilities of the ESB). It is crucial to notice that in the previously mentioned research works, that semantic reconciliation was either manual, either based on specificities of the studied field (see section 5 on crisis management area). This article aims, first at studying how that semantic reconciliation between business models (results from BPM approach) and technical models (deployment files required for workflow orchestration) could be automated, and second at presenting some research results concerning that issue.

The first step is considered as the abstract level while the second one is dedicated to the concrete level. Figure 1 illustrates this global MDE design approach.

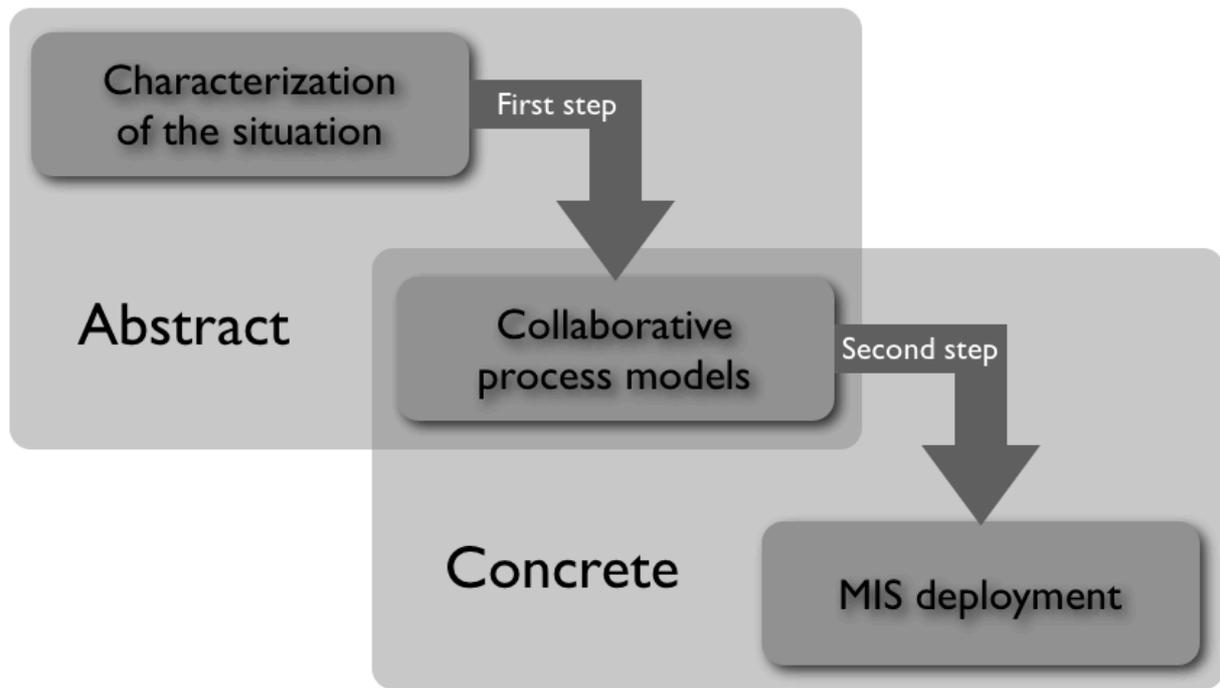

*Figure 1. "Big Picture" of the MISE approach.*

Before detailing this MDE approach step by step, we will focus on the required theoretical elements concerning model transformation principles.

## 2.2 Model transformation principles

Considering the model-driven context of this article, the first crucial point of the presented approach is "model transformation". This issue concerns each of the two previously presented steps (abstract and concrete). The general principle is the following: A source model is used (built according to a source metamodel (MM)) to obtain a target model (respecting a target MM). The key point is that the source MM shares part of its concepts with the target MM. Note that the two spaces, source and target, have to be partially overlapping (on a conceptual point of view) in order to allow model morphism. As a consequence, the source model embeds a shared part (instantiated from the concepts shared by source and target MMs) and a specific part (instantiated from the concepts exclusively contained into the source MM). The shared part provides the extracted knowledge, which may be used for the model transformation, while the specific part should be saved as capitalized knowledge in order not to be lost. Then, mapping rules (built according to the overlapping conceptual area of source and target MMs) can be applied onto the extracted knowledge in order to provide the transformed knowledge. That transformed knowledge and an additional knowledge (to fill the lack of knowledge concerning the instantiation of concepts exclusively contained into the target MM) may be finally used to create respectively the shared part and the specific part of the target model. The way knowledge may be capitalized and/or extracted and/or added may be inspired by model integration principles (Bigand, 2009). This model transformation theoretical framework is synthesized in figure 2.

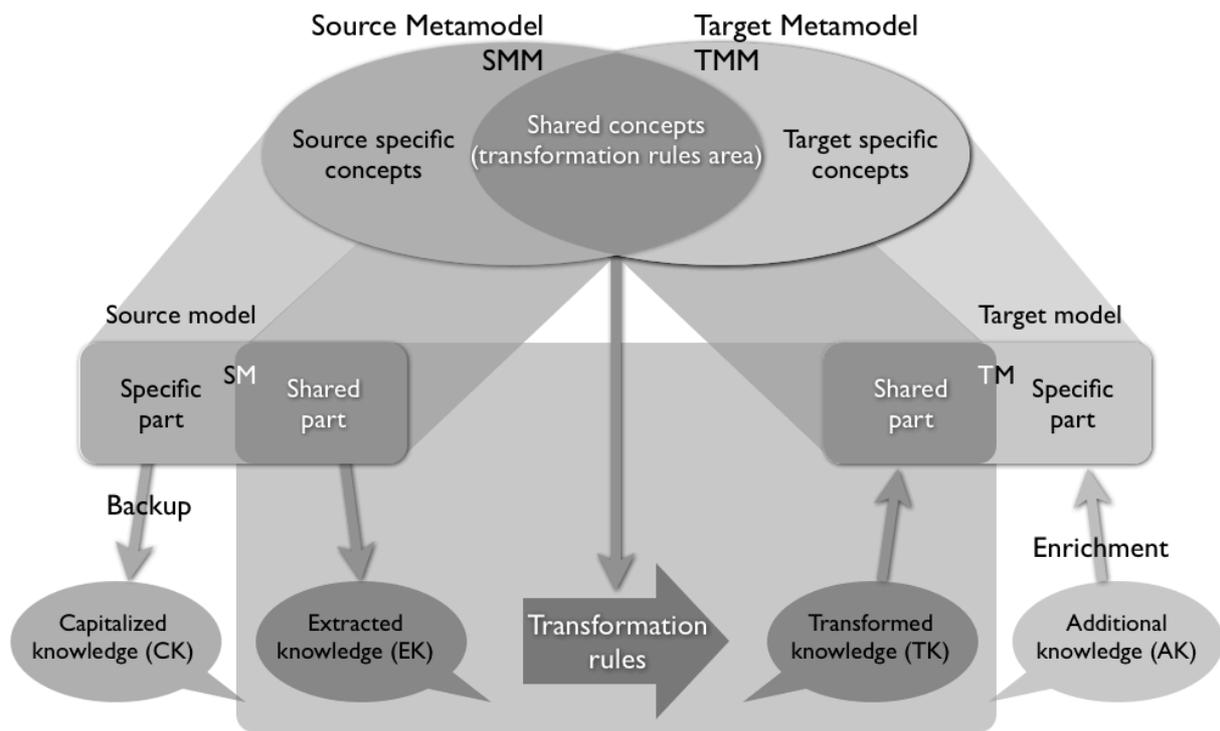

*Figure 2. Model transformation principle.*

Note that both, capitalized knowledge and additional knowledge may be empty depending on the considered model transformation case.

### 2.3    Presentation of the MIS design approach

As exposed before, the MISE design approach includes two steps (abstract and concrete) covering three levels ("characterization of the situation", "collaborative process models" and "MIS deployment"). According to that structure, there are mainly three objectives to achieve: (i) obtaining the first level, (ii) building the transition mechanisms from first level to second level and (iii) building the transition mechanisms from second level to third level.

Achieving first objective requires to build a collaborative situation MM so that the modelling activity might be successful. In the context of MISE, two such MMs have been designed. (Rajsiri, 2009) presents a collaborative situation MM, directly connected to the MIT Process Handbook (Malone, 2003). (Mu, 2011) presents another collaborative situation MM, with strong connections with ISO 9000 principles (especially concerning the types of processes in organisations: decision, operation and support). The main issue of the first goal is to gather the appropriate knowledge concerning a collaborative situation so that the next steps could use it.

Reaching the second target consists in using the collaborative situation model in orders to extract the embedded knowledge and transform it into adequate process cartography. At this stage there are strong semantic issues to align situational elements (objectives, partners roles, abilities, etc.) with process components (activities, flows, events, actors, information, etc.). The main goal is to answer the question "what business activities to support the identified collaborative situation?" and to ensure the semantic alignment between the modelled collaborative situation and a process cartography based on the capacities of partners. To solve that semantic issue, some previous work did use the MIT Process Handbook, which provide matching between the concepts of "objectives" and "processes", as presented in (Rajsiri, 2008) and (Rajsiri, 2009). Some other current research works (Mu, 2011) aims at asking partners to complete semantic description of their competencies in order to ensure the semantic matching more specifically, without being focused on a specific vision of industrial domain (like the MIT Process Handbook). Another solution could consist in hybrid approach, merging semantic description of partners' abilities and some process patterns extracted from different process repositories (MIT Process Handbook, SCORE, etc.). Finally, relatively to section 2.2, the source model is the collaborative situation model (respecting a collaborative situation MM which is the source MM). The additional knowledge is a repository of business activities, containing all the available functions of partners (for instance the MIT Process Hand Book or a any dedicated specific repository). The target MM is a collaborative process MM, for instance the one described in (Touzi, 2009) or the other one described in (Mu, 2011), while the mapping rules may therefore be the ones fully described by (Rajsiri, 2009). There is no specific capitalized knowledge in this model transformation step. The obtained target model is finally a collaborative process model (or a collaborative process cartography, depending on the target MM), dedicated to support dynamically the described collaborative situation.

Concerning the third objective, once the relevant business process cartography has been obtained, the objective is to get a deployable MIS respecting SOA principles. Regarding section 2.2, the source model is the previously obtained collaborative process model, based on the collaborative process MM (which is the source MM). The additional knowledge contains all the technical elements concerning services and data (knowledge embedded into WSDL[1] files). The target MM is the UML technical architecture of the MIS, fully described in (Benaben, 2010) and (Touzi, 2009). (Benaben, 2010) describes the mapping rules. There is no specific capitalized knowledge in this model transformation. The obtained target model is a UML model, describing the technical structure of the MIS (based on the deployment of an ESB). At this stage, there are also strong semantic issues concerning the alignment of the components of business model (processes, activities, information, etc.) and the elements required for the technical deployment on a SOA platform (workflows, services, data, etc.). In MISE, semantic issues of second objective have already been tackled, at least in a first version (Rajsiri, 2009), but, semantic issues of third goal have only be considered so far through a manual point of view without any computerized treatment.

These three objectives and the detailed MISE approach are described in figure 3.

---

[1] Web Service Description Language

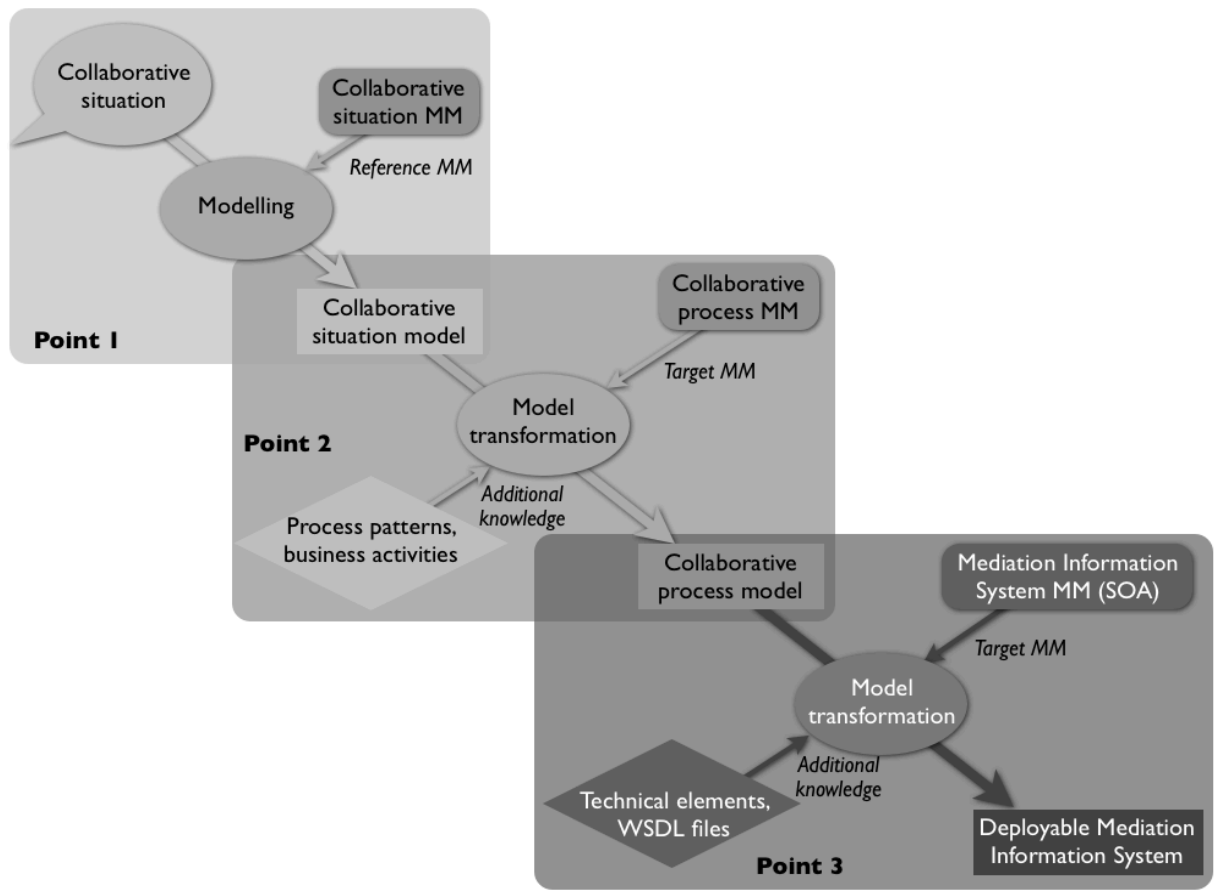

*Figure 3. Detailed model-driven approach of MISE.*

The following section will focus on the semantic issues of the MIS design. In such a design schema, strongly based on model-transformation principles, the semantic gaps are clearly linked to the additional knowledge. Actually, this knowledge brings new elements that have to be identified and correlated to the existing knowledge and the target MM in order to be appropriately integrated in the target model.

### 2.4 Manufacturing application case for MIS platform

In Aerospace industry, the strategy of the main aircraft manufacturers is to outsource more and more sub-assemblies. Relationships between subcontractors, as part of the design and manufacture of composite parts, use different methods of work and are handled through various tools that must be interoperable. The ISTA3 project (for 3rd generation of Interoperability for Aeronautics Sub-contracTors) concerns sub-contractors collaboration in Aerospace industry and focus on information system interoperability as collaboration support. In order to achieve this goal, a collaborative platform was designed, based on MISE architecture and paradigms (BPM, MDI, SOA, etc.).

Our application case treats the collaboration between a composite part manufacturer and its mould producer. It covers the whole collaboration, from pat sketch to mould billing. The collaboration is supported by a mediator, which centralizes the business process and deals with interoperability issues, particularly following semantic ones.

### 3. Semantic issues in MISE project

According to the previous section, the main semantics gaps might be, first at the transition from "collaborative situation model" to "collaborative process model" and second at the transition between "collaborative process model" and "technical MIS deployment model" (two steps where additional knowledge is carrying new semantic knowledge). Only the second semantic issue will be presented and treated in the following. There are three main reasons behind this choice: (i) Both these semantic issues can legitimately be considered as independent. Filling in the first one allows to provide automated BPM approach while filling in the second one allows to computerized any business process. (ii) We believe that the second semantic issue is a very large problem in and of itself. (iii) Finally, some other research works are currently being performed in the same laboratory on the first semantic issue (and will be submitted to publication).

The objective of tackling that semantic issue requires finding technical elements in order to implement strictly business components. Therefore, the global semantic framework of this article is the one presented in figure 4.

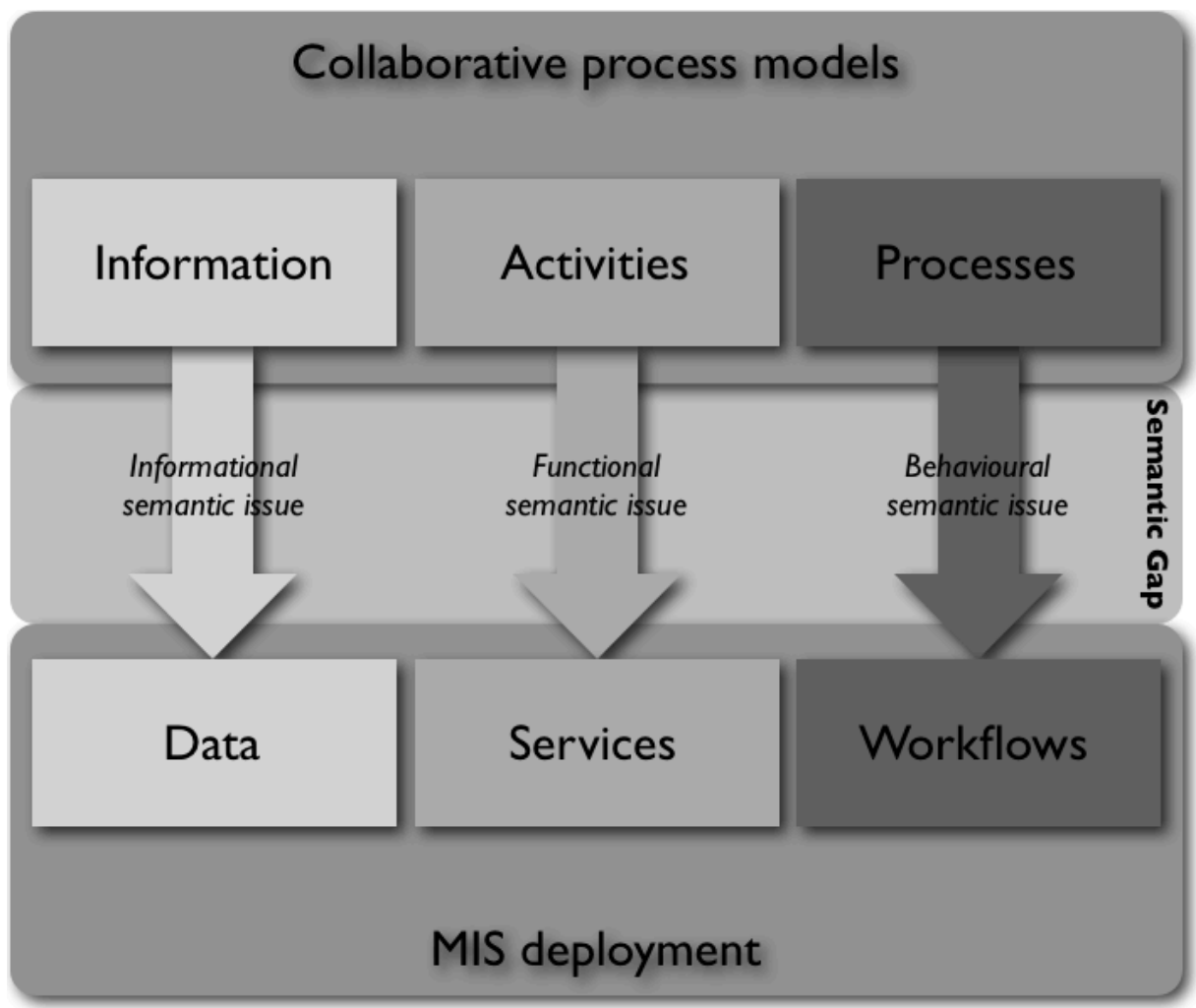

*Figure 4. Semantic issues between business level and technical level.*

Assuming that process models can be obtained (from the previous steps of MISE approach or even from any BPM method), there are clearly three semantic issues, which, once solved, may allow connecting business process to IS deployment in a relevant manner: informational issue, functional issue and behavioural issue. These issues may be formulated as follows: How to deal with information reconciliation? How to ensure the matching between business activities and technical services? How to obtain workflow from business process?

Furthermore, these first two questions deals with many-to-many issues. The objective is to map a set of business activities (respectively information) with a set of technical services (respectively data) during the model transformation (in order to define a technical way to implement precisely a theoretical solution). Besides, information are linked to activities

(activities deal with information as inputs or outputs for example) and data are linked to services (similarly, services deal with data).

Concerning the third question (processes to workflows), it is obvious that the matching between business activities (respectively information) and technical services (respectively data) will not be sufficient to ensure the workflow discovering. There are structural questions to consider: business process includes flows and gateways, which are in charge of process backbone. Once the set of information matched with a set of data and the set of activities matched with a set of services, the question of the way they should be structured and organised is still pending. Of course, this is a direct consequence of the fact that we consider sets of business activities (respectively information) to match with set of technical services (respectively data), in a one-to-one context, the business process structure could be rigorously copy and paste onto the workflow structure: the overall structure of the business process (concerning the way activities are structured in a sequential or parallel manner, with gateways and connections) can be reused as a template to design the overall structure of the technical workflow (by replacing business activities with technical services). That specific point is the final issue to handle in order to manage semantic reconciliation between business level and technical level.

## 4. State of the art concerning semantic reconciliation

A lot of projects or implementations focused on Semantic Web Services (SWS) discovery or composition. SWS are Web Services that are semantically annotated (most of the time, their technical description is extended to embed semantic information). Most of these take interest in one specific SWS representation and are limited to technical service matching.

METEOR-S (Sivashanmugam, 2005) and FUSION (Alexakis, 2007) projects present two approaches of service discovery based on SAWSDL description mapped to an UDDI repository. Whereas the first focused only on a syntactic service matching, (Alexakis, 2007) provides a hybrid semantic matchmaking, using both logic-based reasoning and syntactic similarity measurement. OWLS-MX (Klush, 2006), WSMO-MX (Kaufer, 2006) and SAWSDL-MX2 (Klush, 2009) also provide hybrid semantic matchmaking libraries, respectively for OWL-S, WSMO and SAWSDL. (Klush, 2006) and (Kaufer, 2006) allow the user to manually select a specific similarity metrics while (Klush, 2009) add a Support Vector Machine, which combines variants to improve engine precision.

WSMX (Facca, 2009) is the official WSMO group implementation. It provides a complete execution environment using most of WSMO features such as precondition and effect semantic descriptions. It focuses on a "1-to-1" logic-based service matching thanks to which an abstract process (expressed in WSMO Choreography) is transformed to an executable one. IRS-III (Domingue, 2008) also aims to discover, select and orchestrate web services based on WSMO but adds UPML knowledge model (Unified Problem-solving Method description Language) in order to express process semantic definition. SUPER (Hepp, 2005) is based on IRS-III framework but tries to use semantically annotated BPMN process (called sBPMN), similarly to our objective with BPMN 2.0. It couples pattern reuse and "1-to-n" composition in order to deal with granularity differences between business activities and technical services. Finally, SOA4All (Lécué, 2010) takes interest in large-scale semantic matchmaking and defines a lightweight SWS representation based on WSMO (called WSMO-Lite) and its executive environment based on a light WSMX, improving algorithm performances.

Concerning exchange management between services, (Sivashanmugam, 2005) and (Domingue, 2008) made the choice to give up message transformation and provide graphical interfaces to handle it. (Hepp, 2005) aims at generating necessary transformations using SAWSDL I/O semantic concepts to bind message parts together. Then, it uses ontology axioms to express some expected transformations and let the user complete the transformations through an Eclipse-based GUI. Finally, it fulfils BPEL process with final transformations (thanks to their BPEL4SWS extension) in order to reuse it. (Gagne, 2006) centres on semantic data matching to generate message transformations. It only takes interest in tagging affiliation and does not handle format or value divergences. (Madnick, 2009) focuses on context heterogeneity to transform message values thanks to specific reasoning on semantic format description.

Contrary to (Hepp, 2005), we consider multiple SWS representation during our abstract to concrete transformation. Furthermore, our approach allows "n-to-m" semantic matchmaking, which increase application cases. Finally, the complete integration of abstract and concrete design levels enables business monitoring and runtime workflow adaptation.

## 5. Specific semantic treatment in ISyCri project (MISE 1.0)

The ISyCri project is a French funded project (ANR/06/CSOSG) dealing with Interoperability of Information Systems in Crisis situations. It is mainly focused on providing a crisis management cell with a MIS dedicated to ensure the collaboration between partners of the crisis cell (while these partners are assuming the collaboration with their people on site through their personal and specific channels). To reach that goal, the previously described (section 3) semantic issues had to be solved, in the crisis management context.

### 5.1 Functional semantic issue

We will first explain the functional issue as far as it may be considered as the core-part, essential to understand both the informational issue and the behavioural issue. The solution adopted to deal with that key semantic issue is quite rough but well adapted to the considered field: The added knowledge of the collaborative process modelling step (point 2 of figure 3) is based on a repository of technical services. This is a shortcut to avoid semantic problems at the functional levels but it is quite realistic in crisis management context: Actually, for partners of a crisis management cell, technical services are factual representations of business activities. For example, if policemen are able to establish a safety perimeter on the crisis site (business activity), then, the MIS, which is orchestrating the collaborative workflow inside the crisis management cell, should be able to invoke that action at some point. Therefore, a service should exist in the policemen IS (technical service) in order to be invoked by the MIS. However, that service could not be directly the expected operational service (establishing a safety perimeter is not a computable activity, it is a business activity). This service might be an interface (still a technical service) that informs the policemen delegate inside the crisis management cell (through his IS) that it is time to demand to establish the security perimeter from the policemen on site. This technical service will therefore be a kind of interface, requesting the business activity of establishing a security perimeter. Schematically, in this case, instead of trying to match business activities with technical services, the chosen way oblige to select business activities among technical services (furthermore on a one-to-one schema). The matching is not necessary anymore. This direct way to deal with the functional semantic issue is mainly compliant with the specific context of crisis management (due to the previous consideration on proximity of technical services with business activities) but would not be adapted to other domains where the semantic distance between technical services and business activities would be more consequent.

### 5.2 Informational semantic issue

As for the informational issue, first a part of the required matching is done through the previous mechanism as far as selecting business activities among a repository of technical services implies also to select inputs and outputs of this business activities. Besides, another principle is also in charge of that informational semantic reconciliation: Mediation services (inside the MIS) are in charge of translation and matching between data. If one output information of one business activity is conceptually an input of another business activity, it is necessary that the output of the corresponding technical service is correctly interpreted and used as the input of the following corresponding technical service. Therefore, some static lookup tables (that insure the static matching) have been built, according to the specific considered field. Mediation services have also been built in order to be able to read these tables and to use them in order to transform outputs of technical services into required inputs of other technical services:

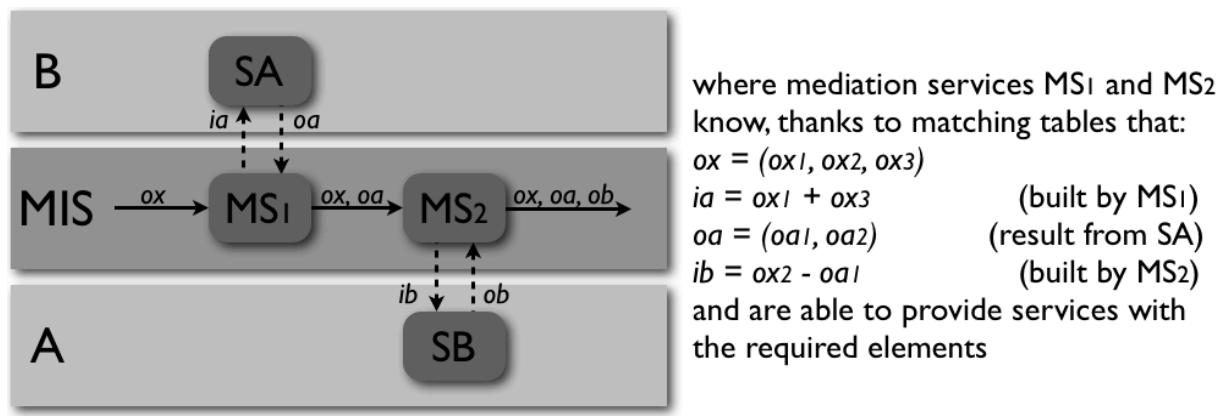

*Figure 5. Mediation services principle.*

The informational semantic issue is then managed by the ability of mediation services to use pre-established semantic lookup tables. Similarly to the functional solution, this is a result specifically dedicated to crisis management context. As a critical point, the design of lookup tables should be automatically managed by knowledge management tools. Currently it is a manual work, thanks to the fact that, in crisis management situation, the existing emergency plans define formally the information required for each activity and where they can be found.

## 5.3    Behavioural semantic issue

Finally, concerning the behavioural issue, there is no real semantic reconciliation due to the fact that the questions introduced in section 3 concerning the structure of the reconstructed workflow is strongly depending on the many-to-many matching. If the semantic reconciliation concerns a set of business activities (respectively information) to map on a set of technical services (respectively data), the question of the structure of the technical workflow to deduce from the one of the business process is crucial, but here, each single business activity (respectively information) is mapped on one single technical service. The exact "flows and gateways" structure of the business process can be copy and paste onto the structure of the technical workflow.

Besides, on a syntactic point of view, the workflow design is based on a BPEL[2] transformation tool (for instance, BPMN 2.0 includes direct mapping rules to BPEL), which proposes a BPEL file, orchestrating technical services and data (already semantically consolidated by the management of informational and functional issue). The behavioural issue is managed through the syntactic mapping (BPEL generation) and the preceding semantic reconciliations (informational and functional).

## 6.    Current research works on semantic reconciliation (MISE 2.0)

Semantic issues have been identified and defined in MIS design context. Furthermore, some specific ways to solve these issues have been presented in the particular field of crisis management. These solutions are not satisfying for sub-contractor collaboration where IS and their data models are usually very different from one partner to another. That is why we are currently working on semantic reconciliation in MISE. SWS precisely aim at the automated discovery, selection and orchestration of Web services on the basis of machine-interpretable semantic descriptions. This allows users to associate semantic concepts to syntactic web service descriptions (WSDL).

On the one hand, we aim at matching business activities and technical services, considering granularity differences between abstract and concrete levels. It involves an "n-to-m" matching during the business to technical transformation and ontology matching of concepts from different levels. On the other hand, we aim at enabling on the fly data translation thanks to automatic transformation services, in order to avoid manual matching. Service and data reconciliations require three phases: (i) knowledge modelling of technical and collaboration domain concepts, (ii) incorporating semantics into business and technical models thanks to existing or new standards, (iii) using semantic information to match services or messages.

---

[2] Business Process Execution Language

### *6.1  From business processes to executable workflows*

In order to execute abstract processes, we have to generate the appropriate BPEL processes. BPMN 2.0 specification already suggests a BPMN to BPEL syntactic mapping. This mapping allows us to transform processes from one MM to another but it does not bring the information required for run time, such as real service endpoints or exact exchanged messages. This information is linked to the semantic gap between the abstract and the concrete levels. In order to achieve this goal, we chose to exploit both the semantic description of the internal behaviour, brought by some SWS representations, and BPMN 2.0 extension mechanism to bring semantic annotations into business activities. It enables operation and input/output (I/O) semantic description as well as internal behaviour in order to perform "n-to-m" matchmaking (such as OWL-S or WSMO-Lite for SWS). Thanks to these internal descriptions, it became possible to match functional concepts (operation description or goal). This graph composition is based on I/O available semantic information in order to ensure model integrity. This whole semantic matching is based on a hybrid approach: Semantic distance between concepts is performed thanks to a logic-based reasoning coupled to a syntactic similarity measurement. This syntactic study uses classical similarity metrics such as Cosine, Extended Jaccard or Jensen-Shannon (Cohen, 2003).

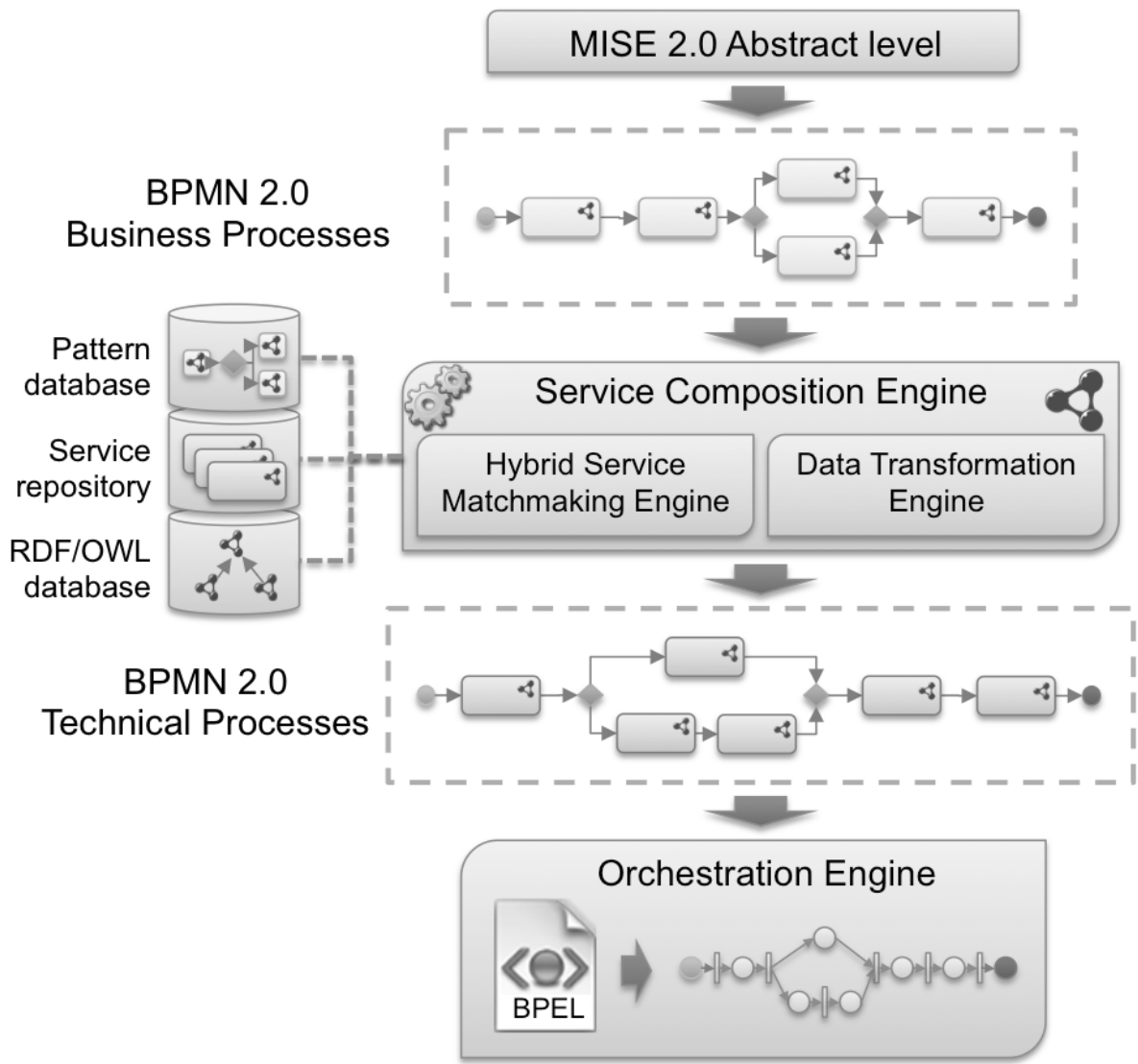

*Figure 6. Overview of business processes to executable workflows transformation.*

The "n-to-m" semantic matchmaking involves a combinatorial computation. For each possible group of activities, we theoretically have to test each possible group of available services. In order to limit possible associations, we consider some simple filters, from both levels (abstract and concrete). (i) Considering process logic, we easily can reduce possible groups. We cannot consider bringing together activities randomly selected. The business process implies connections between activities that we can not avoid (i.e. sequences, gateways). Figure 7 illustrate possible groups of business activities that can be considered for semantic reconciliation, according to possible semantic expression of SWS. (ii) Technical services are also concerned by combinatorial reduction. Some information about collaboration, such as target partner, activity domain or non-functional requirements, could be used as filter. Those filters do not require time consuming computation and can be performed before any semantic matchmaking while this information is contain in our technical registry.

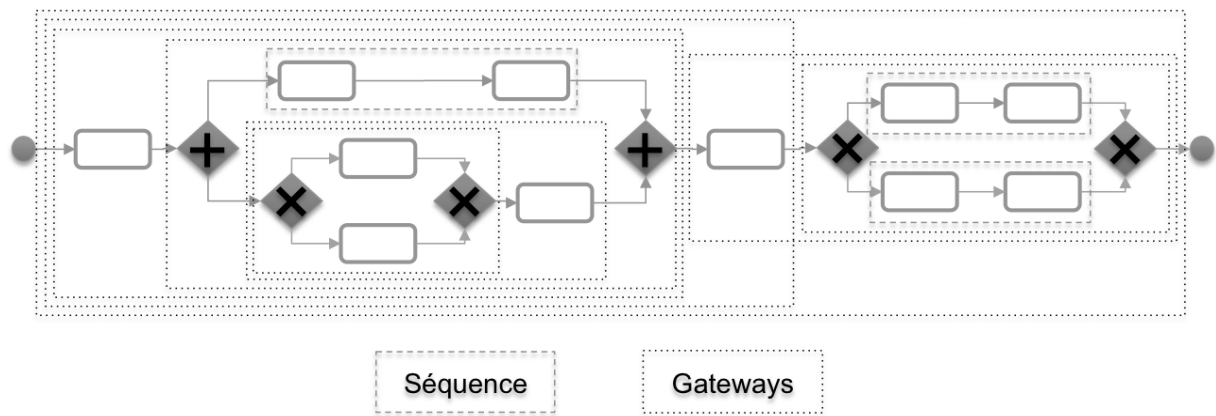

*Figure 7. Possible activity groups according to process logic*

In order to provide reusability and acceptable performances, we have also based our process transformation on a pattern database populated with previous successful tries. The whole process transformation follows those steps: (i) we search for existing patterns in our database; (ii) for uncovered activities, we study semantic description of operations, I/O and internal behaviour in order to get the best technical process possible; (iii) if some activities are still uncovered, we inform the user. He can then choose to develop a new web service, find another partner which already owns it or entrusts our library to generate GUI-based services, which handle expected messages then confide the added value activity to a chosen partner (this is mainly an interface generator).

### *6.2 On the fly data transformation*

The discovery of web services that fit our functional needs is not enough to generate executable processes and insure good communication. We also have to provide interoperability between these services thanks to on the fly data transformation. We selected three steps to apply to each service in order to generate transformation for its inputs. The following approach is performed during the design time but could be resumed at runtime if necessary: (i) we try to match input concepts with previous service outputs thanks to semantic annotations, we could use hybrid matchmaking as in activities' matching. Only few concepts are involved here, so the matchmaking is quite easy. That allows us to found some equivalent tags (without taking interest into data format for now).

Let us take an example and focus on one service input: a sensor recording service which store values from a temperature sensor and return an alert according to business rules. We know it fit with our functional needs according to previous semantic matchmaking. We now have to focus on the technical input matching (XML tags). The Sensor Recording service expects the reading time (Datetime in an SQL Datetime format) and the read value (Fahrenheit). Thanks to semantic annotations, we could use hybrid matchmaking as in activities' matching. Only few concepts are involved here, so the matchmaking is quite easy. We found some equivalent tags from previous values: Date (format US) and Time that cover the whole Datetime embedded concept and SensorTempC which correspond to a temperature in Celsius (see line 1 in figure 8). Then, we look at syntactic format links between tag equivalences. For each input concept and its associated tags, we search in our database for low-level decomposition. (e.g. {#DateUS} = {#Month} – {#Day} – {#Year}). Then, thanks to corresponding regular expression (also in our database) we deduce the format transformation for each input (see line 2). Finally, concerning value divergences we focus on possible unit conversion, based on the same principle than before and coupled with a math expression parser. (e.g. {#SensorTempC} = {#Celcius} = ({#Fahrenheit} x 1,8) + 32, on line 3).

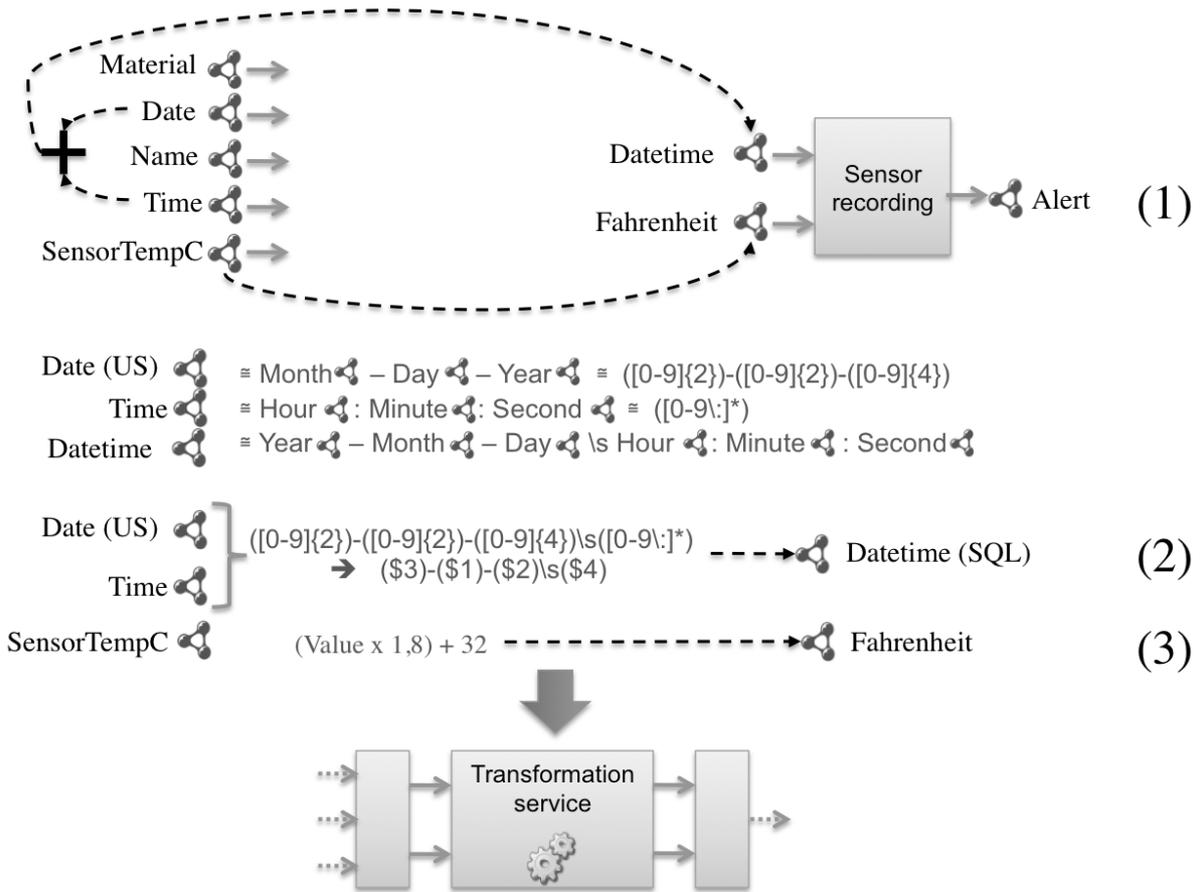

*Figure 8. Semantic data matchmaking in MISE 2.0.*

This generated transformation covers the main format divergences but does not handle data issues. It can be expressed in an XLST file, using only SWS information (semantic message description). In order to manage data conflicts between partners, we add a "1-to-1" syntactic replacement at runtime waiting for a best idea. If one of these steps does not match we have to find another service (which could be considered as a prerequisite of this service for next calls) or entrust the user with the transformation (then fulfil the format database in order to automate similar transformations).

## 7. Conclusion

This article shows the semantic barriers between business models and technical models, in the case of information system design in collaborative context. Besides, this article also aims at positioning these issues in a wider view by showing how such a semantic reconciliation could be part of a larger model-driven approach, which starts from collaborative situation characterization and finishes at the SOA collaborative information system deployment (thanks to a mediation information system in charge of ensuring interoperability between existing information systems). Furthermore, this article also aims at showing how these semantic issues have been specifically tackled in previous research works (MISE 1.0) and how they are currently considered in a more generic vision in MISE 2.0.

Extracted from the MISE context, the presented considerations and results are useful on their own. Business Process Management (BPM) is currently being unavoidable in organisations' life because it is the most powerful way to (i) obtain official certifications (ISO, CMMI, etc.), (ii) improve their behaviour (by studying performance indicators on effectiveness, efficiency and relevance) and (iii) ensure appropriate requirements for information system design or improvement (because IS should respect and support the behaviour of organizations). Thus, building process models and process cartography is a more and more unavoidable stage in the lifecycle of any organization. It should also be the case concerning collaborations and networked organisations. However, considering the third point and assuming the fact that obtaining business process models describing the behaviour of the whole collaboration is nowadays concretely feasible, there is still a deep gap between business process models and information system deployment. Even if service-oriented architectures and workflow-based approaches are slowly filling in that gap, there is still an inescapable semantic difference between business considerations and technical considerations. This article proposes to identify, define and somehow to tackle these semantic issues.

Finally, the whole article is dedicated to design-time: it concerns MIS design from collaborative situation characterization to MIS design through process cartography modelling. However, collaborative situations, even more than single organisation, are strongly expected to evolve and adapt their behaviour to dynamic context. Collaborative situation are supposed to emerge to catch business opportunity and to be very reactive in order to be competitive. These considerations mean that the whole business structure could vary at any time, involving strong changes also at the technical level. Semantic reconciliation should so be used also when on-the-fly reconfiguration is required. By providing semantic reconciliation tools as web-services, it is possible to define MIS design-workflows, so that when adaptation is required, it might be possible to obtain new business models and to transform them into new technical models. These issues are currently being considered through some research works on event-based architectures (EDA). Actually, two research projects (National and European, started in 2010) and two PhD (started in October 2010 and October 2011) are supporting these current and future research works on automatic management of events (publish/subscribe mechanism for web services). This article shows that one ambition of MISE project is to provide automation in design-time of IS mediation. The next ambition, thanks to EDA, is to provide automation in run-time of IS mediation. These future works aims at supporting the functions of gathering, storing, filtering, analysing, combining and exploiting computerized events (emitted by any services). This event management can be considered as a formal monitoring of the collaborative behaviour and will be used to support the MIS agility by keeping it as relevant as possible to a collaborative situation that probably evolves.